\documentclass[mathleft
]{an}
\usepackage{graphicx}
\usepackage{times}
\begin{document}

\Pagespan{789}{}
\Yearpublication{2006}%
\Yearsubmission{2005}%
\Month{11}%
\Volume{999}%
\Issue{88}%

\title{Accreting White Dwarfs As Supersoft X-ray Sources }

\author{M. Kato\inst{1}\fnmsep\thanks{Corresponding author:
  \email{mariko@educ.cc.keio.ac.jp}\newline}
}
\authorrunning{M. Kato}
\institute{
Keio University, Hiyoshi, Kouhoku-ku, Yokohama, 223-8521, Japan
}

\received{30 July 2009}
\accepted{11 Nov 2009}
\publonline{later}


\abstract{
I review various phenomena associated with mass-accreting white dwarfs (WDs) 
in the view of supersoft X-ray sources.
When the mass-accretion rate is low 
($\dot M_{\rm acc} < $ a few $\times 10^{-7} \mathrm{M_\odot}$yr$^{-1}$), 
hydrogen nuclear burning is unstable and nova outbursts occur.  
A nova is a transient supersoft X-ray source (SSS) in its later phase 
which timescale depends strongly on the WD mass. The X-ray turn
on/off time is a good indicator of the WD mass. 
At an intermediate mass-accretion rate 
an accreting WD becomes a persistent SSS with steady hydrogen burning.
For a higher mass-accretion rate, the WD undergoes ``accretion wind evolution''
in which the WD accretes matter from the equatorial plane and loses mass 
by optically thick winds from the other directions. 
Two SSS, namely RX\,J0513$-$69 and V\,Sge, are corresponding objects to 
this accretion wind evolution. 
%
%
We can specify mass increasing WDs from light-curve analysis based on 
the optically thick wind theory using 
multiwavelength observational data including optical, IR, and supersoft X-rays. 
Mass estimates of individual objects give important information 
for the binary evolution scenario of type Ia supernovae.
	}

\maketitle

\section{Introduction}

Accreting white dwarfs (WDs) become transient, intermittent, and persistent 
supersoft X-ray sources (SSSs) depending on the mass-accretion rate. We find 
various phenomena for wide ranges of time-scales and wavelength.  
In this paper I will review when and how the supersoft X-rays emerge from 
the WDs. 
Section 2 briefly introduces the stability a-nalysis of accreting WDs. 
Section 3 deals with low mass-accretion rates; WDs experience nova outbursts 
and become transient SSSs in the later phase. 
With intermediate accretion rate hydrogen burning is stable and 
the WDs become persistent SSSs, which is the subject of Section 4.   
In case of high mass-accretion rate, optically thick wind inevitably occurs 
from the WD surface (accretion wind). 
Section 5 introduces quasi-periodic SSSs as a related object to this regime.

\section{Stability analysis of accreting WDs}
Sienkiewicz (1980) examined thermal stability 
of steady-state models for accreting WDs of various mass.
For low accretion rate, the envelope is thermally unstable which 
triggers a hydrogen shell flash, but for higher accretion rate, nuclear 
burning is stable. Nomoto et al. (2007) reexamined this stability using 
OPAL opacity and confirmed Sienkiewicz' results. The stable and unstable
regions are denoted in Figure \ref{hr_simple}. 
The unstable part represents WDs with thin envelope
in which energy generation is mainly due to compressional heating, and stable region does 
WDs with nuclear burning at the bottom of an extended envelope.

 

\begin{quote} 
\begin{figure}
\includegraphics[width=80mm]{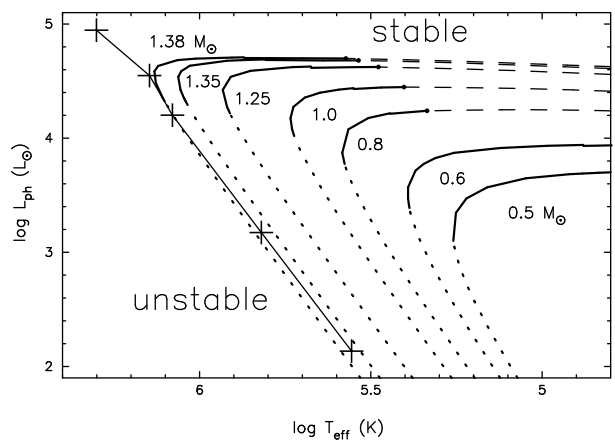}
\caption{Loci of accreting WDs in the HR diagram.
Each sequence corresponds to a mass of the WD which
accretes matter of solar composition.  
The dotted part indicates unstable hydrogen burning, and solid and 
dashed parts stable burning. 
The dashed part is the region of optically thick wind mass loss in 
which supersoft X-ray flux cannot be expected due to the self absorption 
by the wind.  
``$+$'' marks connected with a solid line denote ``stable'' solution  
claimed by Starrfield et al. (2004) for 1.35 $\mathrm{M_\odot}$.}
\label{hr_simple}
\end{figure}
\end{quote}

These results on stability for steady-state models are consistent with
evolutionary calculations of hydrogen-shell flashes on accreting WDs
(e.g., Paczy\'nski \& \.Zytkow 1978; Sion et al. 1979; Prialnik \&  Kovetz 1995;
Sparks, Starrfield \& Truran 1978; Nariai, Nomoto \& Sugimoto 1980; 
Townsley \& Bildsten 2004) 
and also with static envelope analysis (Iben 1982, and Sala \& Hernanz 2005).
Therefore, these results on stability of accreting WDs are considered
as a kind of consensus among the researchers in this field.

It has to be noticed that Starrfield et al. (2004) presented different results 
for accreting WDs, that  they call ``surface hydrogen burning models''. In their 
calculation, WDs stably burn hydrogen for the accretion rates ranging 
from $1.6\times 10^{-9}$ to $8.0\times 10^{-7}~\mathrm{M_\odot}$yr$^{-1}$ and become
type Ia supernovae.  
This result is in contradiction to our present understandings of stability of shell flash 
and all of the previous numerical results cited above.
Nomoto et al. (2007) pointed out that these stable ``surface hydrogen burning''
is an artifact which arose from the lack of resolution in the envelope 
structure of Starrfield et al.'s models.

\begin{quote} 
\begin{figure}
\includegraphics[width=80mm]{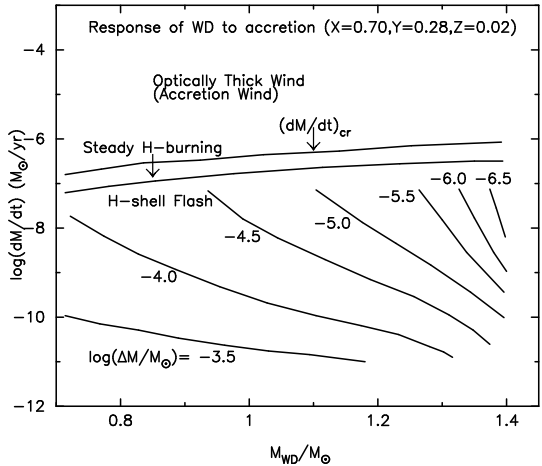}
\caption{Response of WDs to mass accretion is illustrated
in the WD mass and the mass-accretion rate plane.
In the region above $\dot M_{\rm cr}$ strong optically thick winds blow.
Hydrogen shell burning is stable for the region of 
$\dot M_{\rm acc} > \dot M_{\rm std}$.
Steady hydrogen shell burning with no optically thick winds occur 
between the two horizontal lines, i.e., 
$\dot M_{\rm std} \le \dot M_{\rm acc} \le \dot M_{\rm cr}$.
There is no steady state burning at $\dot M_{\rm acc} < \dot M_{\rm std}$, 
where unstable shell flash triggers 
nova outbursts. The ignition mass for shell flash is indicated beside the 
locus of the same ignition mass. 
See Hachisu and Kato (2001) for more detail.
}
\label{accmap_z02}
\end{figure}
\end{quote}
Figure \ref{accmap_z02} shows the response of the accreting WDs 
in the mass-accretion rate vs. WD mass diagram.  
The lower horizontal line denotes $\dot M_{\rm std}$, the boundary 
between the stable and unstable regions of nuclear 
 burning (i.e., the boundary of solid and dotted 
regions in Figure \ref{hr_simple}). 
If $\dot M_{\rm acc} < \dot M_{\rm std}$, hydrogen shell burning is unstable
and nova outbursts occur.  Otherwise, hydrogen burning is stable and no
nova occurs.

The upper horizontal line in Figure \ref{accmap_z02} indicates $\dot M_{\rm cr}$, 
the boundary that the optically thick winds occurs (i.e., the small circle 
at the left edge of the dashed line in Figure \ref{hr_simple}.)
In the region above $\dot M_{\rm cr}$, 
strong optically thick winds always blow (Kato and Hachisu 1994, 2009).

In the intermediate accretion rate, i.e., 
$\dot M_{\rm std} \le \dot M_{\rm acc} \le \dot M_{\rm cr}$,
the hydrogen burning is stable and no wind mass loss occurs. The WD burns
hydrogen at the rate equal to mass accretion, and the WD keeps staying on 
the same position on the thick part in the HR diagram. 
The surface temperature is high enough to emit supersoft X-rays 
(see Figure \ref{hr_simple}).
Therefore, this region corresponds to persistent X-ray sources.

\begin{quote} 
\begin{figure}
\includegraphics[width=80mm]{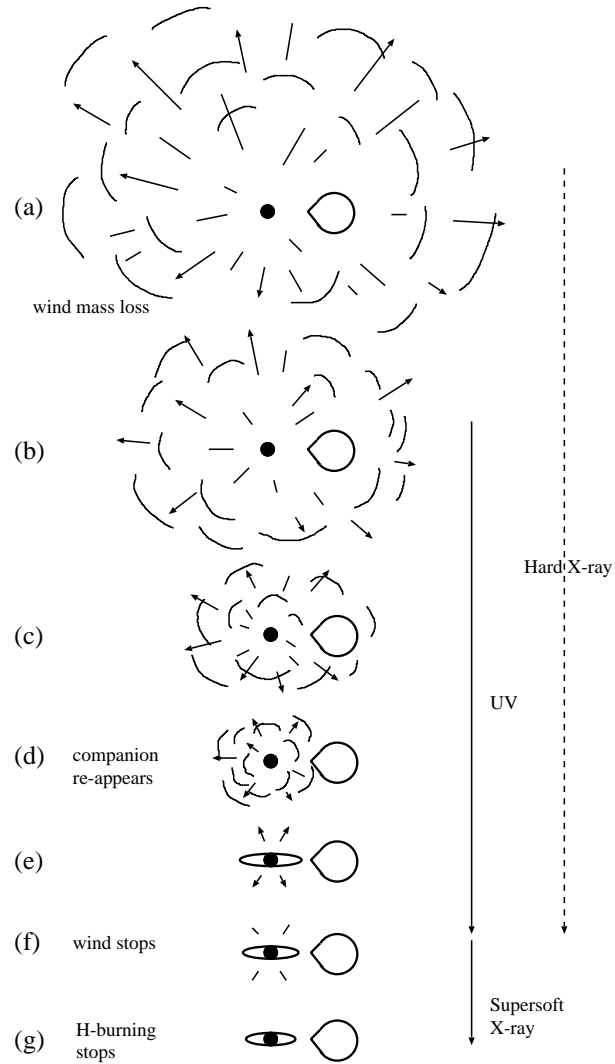}
\caption{Evolution of nova outbursts. After the nova explosion sets in,
the companion star is engulfed deep inside the photosphere (a);
the photospheric radius moves inward with time due to strong mass 
loss. The companion  emerges from the WD photosphere (d) and
an accretion disk may appear or reestablish again (e). 
The optically thick wind stops (f). 
Hydrogen nuclear burning stops and the nova enters a cooling phase (g).
The main emitting wavelength region shifts from optical to UV and then 
to supersoft 
X-rays. (taken Hachisu and Kato 2006) 
}
\label{novaexplosion}
\end{figure}
\end{quote}

\section{Low mass-accretion rate}

\subsection{Nova as transient SSSs}

When the mass-accretion rate onto a WD is smaller than the critical 
value ($\dot M_{\rm acc} < \dot M_{\rm std}$),  
unstable hydrogen shell flash triggers a nova outburst. The WD envelope quickly 
expands and it moves from the lower region (Fig. 1, dotted region) to the 
upper right region in the HR diagram. 

Figure \ref{novaexplosion} shows the evolutional change of nova binary during an
outburst.  
After the nova outburst sets in, the envelope of the WD widely expands 
and strong wind mass loss begins. The optical photons dominate 
in the first stage which 
is replaced by the UV and then the X-ray photons as the photospheric temperature 
rises with time. The time scale of optical decline, UV and X-ray phases 
depends strongly on the WD mass and secondary
on the chemical composition (e.g. Hachisu and Kato 2006).  
In general, a nova on a massive WD evolves fast so duration of the X-ray 
phase is also short, but for less massive WDs it lasts long. 
From the theoretical point of view, all novae become SSS in the later phase of 
the outburst, although the time scale is very different from nova to nova.   

Supersoft X-rays are probably observed only after the optically thick
wind stops, because supersoft X-rays are
absorbed by the wind itself. Therefore, the X-ray turn on time and turn off 
time correspond to the epoch when the wind stops (f) and when hydrogen burning 
stops (g), respectively, in Figure \ref{novaexplosion}. 

 Hard X-rays originate from internal shocks between ejecta (Friedjung 1987; 
Cassatella et al. 2004; Mukai and Ishida 2001) 
or between ejecta and the companion (Hachisu and Kato 2009b), therefore, it can  
be detected during the period as indicated by the dashed line in Figure \ref{novaexplosion}.
%
%
\begin{quote} 
\begin{figure}
\includegraphics[width=80mm]{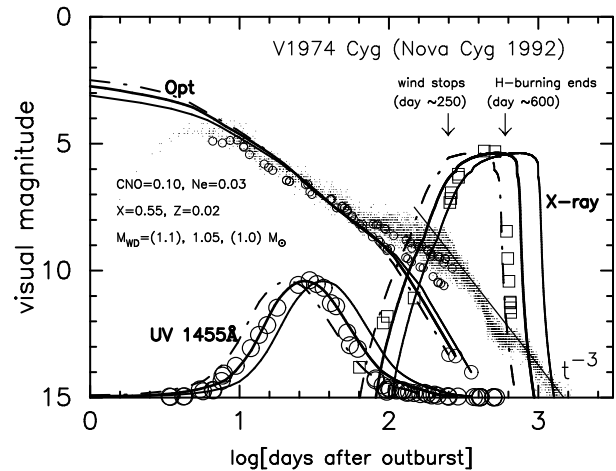}
\caption{Light-curve fitting for V1974 Cyg.  
The supersoft X-ray data (open squares) as well as the UV 1455 \AA~ 
(large open circles), visual (small dot) and $V$-magnitudes (small open circle)
are shown.
The lines denote theoretical curves for a chemical composition of $X= 0.55$,
$X_{\rm CNO}= 0.10$, $X_{\rm Ne}= 0.03$, and $Z= 0.02$.
The model of $1.05 \mathrm{M_\odot}$ WD (thick solid line) shows a best fitting to
these observational data simultaneously. 
Two epochs, which are observationally suggested, are indicated by an arrow:
 when the optically thick wind stops and when the hydrogen shell-burning ends.
(taken from Hachisu and Kato 2006)}
\label{V1974Cyg}
\end{figure}
\end{quote}

\subsection{Light-curve fitting of classical nova}
%
%
\begin{quote}
\begin{figure}
\includegraphics[width=80mm]{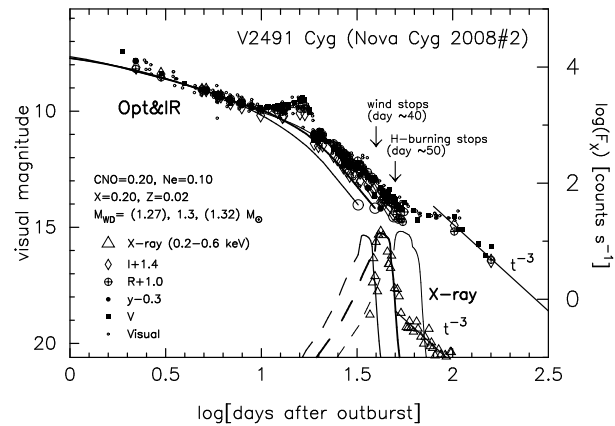}
\caption{Light-curve fitting for V2491 Cyg. 
The upper bunch of data indicates optical and near-IR observational data, and 
the lower X-ray data. 
The best-fit theoretical model is a $1.3~\mathrm{M_\odot}$ (thick blue line)
for the envelope chemical composition with 
$X=0.20$, $Y=0.48$, $X_{\rm CNO} =0.20$, $X_{\rm Ne} =0.10$, and $Z=0.02$.
Supersoft X-rays are probably not detected
during the wind phase (dashed part) because of self-absorption by the wind itself.
The $F_\lambda \propto t^{-3}$ law is added for the
nebular phase. See Hachisu and Kato (2009a) for more detail.
}
\label{V2491Cyg}
\end{figure}
\end{quote}
Nova light curves can be theoretically calculated using optically thick 
wind theory of nova outburst for a given 
set of WD mass and chemical composition of the envelope  (Kato and Hachisu 1994). 
In general, novae evolve fast in massive WDs and slowly in less massive 
WDs, mainly due to the difference of ignition mass (less massive 
ignition mass in massive WDs). The optical and 
infrared (IR) fluxes can be basically well represented by free-free emission. 
There found a beautiful scaling law of optical and IR fluxes among 
a number of novae in different speed class, i.e., 
"universal decline law of classical nova" (Hachisu and Kato 2006). 
This property is useful to 
understand nova light curves that show a wide range of varieties. For 
example, we can extract a basic shape from a given light curve  
and recognize secondary shapes such as oscillatory behavior, 
multiple peaks, sudden optical drop associated to dust formation, and 
additional brightness due to emission lines in the nebula phase.

Figure \ref{V1974Cyg} shows an example of light-curve fitting.
The lines marked ``opt'' represent calculated light curves. In the 
later phase the visual light curve deviates from the theoretical lines due to 
contribution of strong emission lines (see Hachisu and Kato 2006 for more detail). 

The decline rate of optical flux and durations of UV and X-ray 
fluxes depend 
differently on the WD mass and composition, therefore, multiwavelength observation
is important to determine these parameters.  In this case, the above authors 
determined 
the WD mass to be about 1.05 $\mathrm{M_\odot}$ for a set of chemical composition shown 
in the figure caption.

The second example of light-curve fitting is V2491 Cyg. This nova is a 
very fast nova of which supersoft X-ray phase lasts only 10 days. 
Figure \ref{V2491Cyg} shows the best fit model, that reproduces simultaneously   
the light curves of visual, IR and X-ray, is  
$\approx\,1.3\,\mathrm{M_\odot}$ WD with the set of chemical composition 
given in the legend of the figure. This nova shows the secondary maximum 
about 15 days from the optical peak. Except this secondary maximum and 
the very later nebula phase the optical and IR light curves follow the 
universal decline law which is indicated by solid lines. 
(see Hachisu and Kato 2009a 
for the magnetic origin of the secondary maximum.) 
\subsection{X-ray turn on/off time and WD mass}
Hachisu and Kato (2009b) presented light-curve analysis for more than ten 
novae in which supersoft X-rays are detected and determined the WD mass.  
For example, $0.85~\mathrm{M_\odot}$ for V2467 Cyg (CO nova), $0.95~\mathrm{M_\odot}$ 
for V458 Vul (CO nova),  $1.15~\mathrm{M_\odot}$ for V4743 Sgr, and 
$1.2~\mathrm{M_\odot}$ for V597 Pup.

Kato, Hachisu and Cassatella (2009) suggested that  Ne novae   
have a more massive WD than CO novae and the boundary of CO and Ne 
WDs is at $\approx\,1.0\,\mathrm{M_\odot}$ from their mass estimates for seven IUE novae. 
The mass estimates in X-ray nova (Hachisu and Kato 2009b) is consistent 
with the above boundary of $\approx\,1.0\,\mathrm{M_\odot}$  
 although the chemical composition is not known in some novae.
Umeda et al. (1999) obtained  
that the lowest mass of an ONeMg WD is $1.08~\mathrm{M_\odot}$  
from evolutional calculation of intermediate stars in binary. 
This means that a WD is not eroded much, even though 
it had suffered many cycles of nova outbursts.
This may provide interesting information for binary evolution scenarios 
and chemical evolution of galaxies.

\subsection{Recurrent novae}
Recurrent novae repeat outbursts every 10-80 years. The evolution of the 
outburst is very fast. As the heavy element 
enhancement is not detected, their WD mass is supposed to  
increase after each 
outburst. One of the interesting light curve properties is the presence of 
plateau phase.  U Sco shows a plateau phase of 18 days (Hachisu et al. 2000)
and RS Oph 60 days which are an indication 
of the irradiated disk (Hachisu et al. 2006). Hachisu, Kato and Luna (2007) showed 
that the turn off epoch of supersoft X-ray corresponds to the sharp drop 
immediately after the optical plateau phase (see Figure \ref{V2491Cyg.RSOph}); 
They presented an idea that the long duration of the plateau 
in RS Oph is a results of additional heat flux from hot helium ash layer 
developed underneath the hydrogen burning zone.
Therefore, the plateau is another evidence of increasing WD mass. 
%
\begin{quote} 
\begin{figure}
\includegraphics[width=65mm]{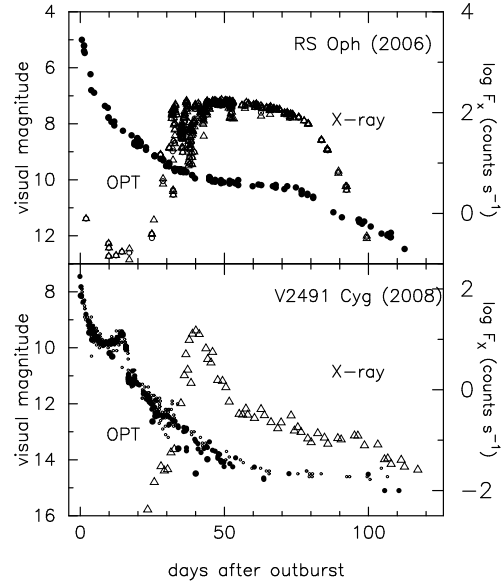}
\caption{Comparison of light curves of RS Oph and V2491 Cyg.  X-ray
 count rates and optical magnitudes are denoted by open triangles and 
filled circles, respectively. RS Oph data is taken from Hachisu et al. (2007), and 
V2491 Cyg data from Hachisu (2009a).
 }
\label{V2491Cyg.RSOph}
\end{figure}
\end{quote}
It is interesting to compare the visual and X-ray light curves of RS Oph 
with a classical nova V2491 Cyg. These objects show a similar rapid 
decline in the first optical phase except the secondary maximum of V2491 Cyg, 
and contain a very massive WD ($1.35~\mathrm{M_\odot}$ in RS Oph: 
Hachisu et al. 2007 and         $1.3~\mathrm{M_\odot}$ in V2491 Cyg). 
However, RS Oph shows a long duration of supersoft X-ray phase, while V2491 does not.

This difference may be explained by the presence of a hot ash layer. 
In classical novae, hydrogen ignites somewhat below the WD surface due to diffusion 
during the long quiescent phase (Prialnik 1986), and ash produced in 
nuclear burning is carried upwards by convection and blown off in the winds. 
Then no helium layer develops underneath the burning zone. Heavy element 
enrichment observed in ejecta may support this hypothesis. 
On the other hand, in recurrent novae, diffusion process
does not work in a short quiescent period, so hot helium ash can pile up 
and act as heat reservoir. This hypothesis needs to be examined more, 
perhaps in a next recurrent nova outburst. 

\section{Intermediate mass-accretion rate}
In the intermediate mass-accretion rate
($\dot M_{\rm std} \le \dot M_{\rm acc} \le \dot M_{\rm cr}$), 
the hydrogen burning is stable and optically thick winds do not occur.
The photospheric temperature of the WD is relatively high as indicated 
by solid lines in Figure \ref{hr_simple}. These WDs are observed 
as persistent SSSs.

\subsection{Steady hydrogen burning}

van den Heuvel et al. (1992) interpolated supersoft X-ray sources as 
an accreting WD with high accretion rate ($\approx\,10^{-7}\,\mathrm{M_\odot}~$yr$^{-1}$) 
so that it can undergo steady hydrogen nuclear burning.  
Figure \ref{SSX.SMC13} indicates the position of the SSSs in the HR diagram 
which are roughly consistent with theoretical steady burning phase 
(thick part), considering difficulties in determining observationally  
the temperature and luminosity.
%
\begin{quote}
\begin{figure}
\includegraphics[width=80mm]{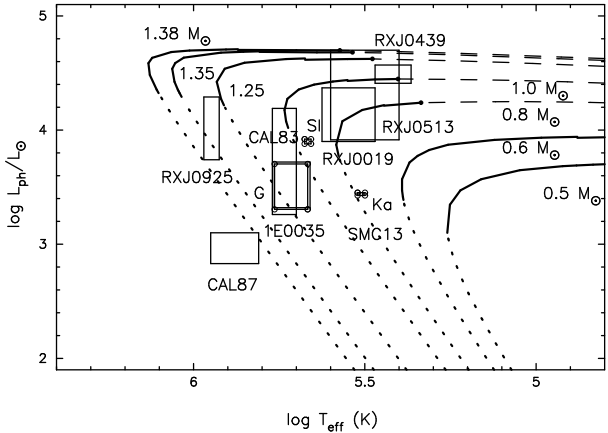}
\caption{Same as Figure \ref{hr_simple} but with SSSs 
(taken from Starrfield et al. 2004 except 1E0035).
Three squares with small open circles denote SMC13 (G: Greiner (2000),  
SI: Suleimanov and Ibragimov (2003), K:Kahabka et al. (1999).
}
\label{SSX.SMC13}
\end{figure}
\end{quote}
\subsection{SMC13: a possible very slow nova?}

It is to be noticed that some supersoft X-ray sources may 
be not exactly steady burning sources, but may be a remnant of 
nova outburst of very slow evolution. 
Kahabka and Ergma (1997) proposed an idea that the observational data of 
1E0035.4-7230 (SMC13) can be explained in the framework of standard 
cataclysmic variable evolution of low mass WDs ($\approx$ 0.6-0.7 
$\mathrm{M_\odot}$). 

Figure \ref{lightM04.Z} demonstrates that a low mass WD (0.4 $\mathrm{M_\odot}$) 
undergoes nova outburst of extremely slow evolution.
Its X-ray turn on/off times are 300 and 600 yrs, respectively for $Z=0.02$
 and more slower for population II stars ($Z=0.004$ and 0.001). 
In these cases the supersoft X-ray phase starts when the  
optical magnitude drops by 6 mag, long after the optical peak. Therefore, 
we can detect no optical counter part of a SSS nor find any 
record in literature.

Figure \ref{SSX.SMC13} also shows the estimated position of SMC13 by three squares  
with small open circles at each corners (two squares are very small). 
These positions are scattered among authors with different method of analysis,   
but roughly consistent with solid part of the theoretical lines (a persistent 
source). If SMC13 is a very slow nova, its X-ray emission more than 
a decade suggests a less massive WD ($<0.6~\mathrm{M_\odot}$) that has a smaller 
temperature. Thus, it is valuable to update the temperature and luminosity 
of SMC13 using  
unanalyzed high quality data recently obtained with satellites after BeppoSax 
and ROSAT.
%
%
\begin{quote} 
\begin{figure}
\includegraphics[width=80mm]{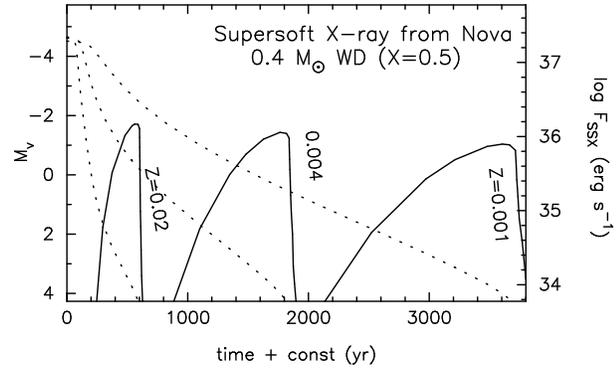}
\caption{Theoretical light curves of visual and supersoft X-ray (0.1-0.6 keV) 
fluxes for $0.4~\mathrm{M_\odot}$ WD of various population ($Z=0.02,~0.004$ and 0.001).
}
\label{lightM04.Z}
\end{figure}
\end{quote}
\section{High mass-accretion rate}
\subsection{Accretion wind}

When the accretion rate is larger than $\dot M_{\rm cr}$, 
the WDs cannot consume all of the accreted matter which is piled
up to form an extended envelope. As the 
photospheric temperature decreases to reach the critical value 
(i.e., the rightmost point of the thick part in 
Figure \ref{hr_simple}), optically thick winds is accelerated due to 
Fe peak (at around $\log T ($K$)\approx 5.2$) of the 
OPAL opacity (Kato and Hachisu 1994, 2009).
 
Hachisu and Kato (2001) proposed a binary system in which the WD accretes 
matter from the companion from the equatorial region and loses matter as 
a wind from the other regions as illustrated in Figure \ref{accwind}. 
They named such a configuration ``accretion wind''.

In such a case, the WD burns hydrogen at the rate of $\dot M_{\rm nuc}$ and 
blows the rest of the accreted matter in the winds at the rate of about 
$\dot M_{\rm acc} - \dot M_{\rm nuc}$, 
where $\dot M_{\rm nuc}$ is the nuclear burning rate.  
Such a WD in this "accretion wind"  
corresponds to the dashed part in Figure \ref{hr_simple}.

This accretion wind is an important elementary process for binary evolution 
scenario to Type Ia supernova, because it governs the growth rate of the WD mass 
(e.g., Hachisu, Kato \& Nomoto 1999a, Hachisu et al., 1999b, Han \&
Podsiadlowski 2006),  
as well as the mass-transfer rate from the companion which is regulated 
by stripping of companion surface by the wind 
(e.g., Hachisu, Kato \& Nomoto 2008). 
%
%
\begin{quote} 
\begin{figure}
\includegraphics[width=65mm]{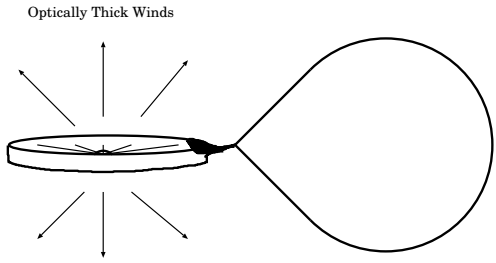}
\caption{Optically thick winds blow from mass-accreting WDs 
when the mass-transfer rate from a lobe-filling companion exceeds a critical 
rate, i.e., $\dot M_{\rm acc} > \dot M_{\rm cr}$.  
The white dwarf accretes mass from the equatorial region and 
at the same time blows winds from the polar regions.}
\label{accwind}
\end{figure}
\end{quote}
%
%
%
\begin{quote}
\begin{figure}
\includegraphics[width=80mm]{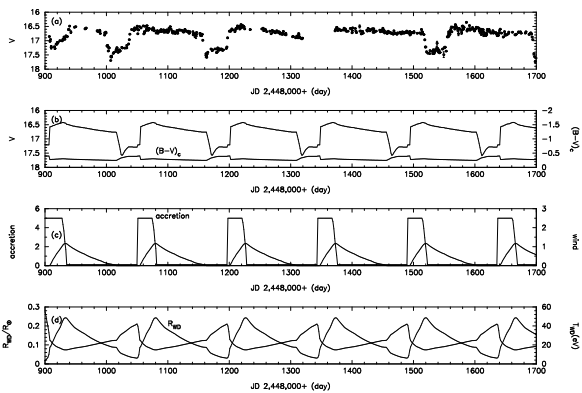}
\caption{Self-sustained model of spontaneous winds for RX J05134$-$6951. 
(a) long term evolution of V magnitude. 
(b) model light curve of $M_{\rm WD}=1.3~\mathrm{M_\odot}$. 
(c) change of accretion rate and wind mass-loss from WD envelope. 
(d) Change of WD radius and its temperature. 
(Taken from Hachisu and Kato 2003b.) 
}
\label{rxj0513}
\end{figure}
\end{quote}
\subsection{Accretion wind and SSS}

There are two objects closely related to the accretion winds: RX J0513$-$69 and 
V Sge. Both of them are supersoft X-ray sources. 

RX J0513$-$69 is an LMC SSS that shows quasi-regular transition between 
optical high and low states as shown in Figure \ref{rxj0513} in which 
supersoft X-rays are detected only in the optical low states (Reinsch et al. 2000; 
Schaeidt, Hasinger, and Truemper 1993).
 
Hachisu and Kato (2003b) presented a transition mechanism between the high 
and low states. In the optical high state, the accretion rate is high enough  
and the photosphere expands to accelerate the winds (Figure \ref{accwind}). The WD 
locates in the low temperature region (dashed part in Figure \ref{SSX.SMC13}) 
and no X-rays are expected. 
In the optical low state, mass-accretion rate is low and 
the photospheric temperature is enough high 
(in the solid part of Figure \ref{SSX.SMC13}) to emit supersoft X-rays. 
No wind is accelerated. 
The above authors proposed a self-regulation transition 
mechanism that makes the binary 
back and forth between the optical high and low states.
When the mass-accretion rate is large, the WD is in the optical 
high state. The strong winds hit the companion and strip off 
a part of the companion surface. Thus the mass-transfer rate 
onto the WD reduces and finally stops, which causes the wind stop and 
the system goes into the optical low state. After a certain time, the 
companion recovers to fill the Roche lobe again and the mass transfer resumes, 
which cases wind mass loss. 
The resultant theoretical light curves depend on the WD mass and other 
parameters. The best fit model that reproduces 
the observed light curve best indicates the WD 
mass to be 1.2 - 1.3 $\mathrm{M_\odot}$ (see Figure \ref{rxj0513}).

The second object is V Sge that also shows the similar semi-regular transition 
of light curve, although timescales are different. 
Its light curve is also reproduced by the transition model with 
the WD mass of 1.2 - 1.3 $\mathrm{M_\odot}$ (Hachisu and Kato 2003a).

In these two systems the WD mass is increasing with time, because steady 
nuclear burning produces helium ash which accumulate on the WD. Therefore,   
they are candidates of type Ia supernova progenitor.




\end{document}